\documentclass[twocolumn]{aastex631}

\begin{document}

\title{Destratification in the Progenitor Interior of the Mg-rich Supernova Remnant N49B}

\author[/0000-0001-9267-1693]{Toshiki Sato}
\affil{Department of Physics, School of Science and Technology, Meiji University, 1-1-1 Higashi Mita, Tama-ku, Kawasaki, Kanagawa 214-8571, Japan}

\author[0009-0003-0653-2913]{Kai Matsunaga}
\affil{%
 Department of Physics, Graduate School of Science, Kyoto University, Kitashirakawa Oiwake-cho, Sakyo-ku, Kyoto 606-8502, Japan
}%

\author[0000-0003-4876-5996]{Ryo Sawada}
\affil{%
 Department of Earth Science and Astronomy, Graduate School of Arts and Sciences, The University of Tokyo, Tokyo 153-8902, Japan
}%

\author[0000-0002-6705-6303]{Koh Takahashi}
\affil{%
 National Astronomical Observatory of Japan, National Institutes for Natural Science, 2-21-1 Osawa, Mitaka, Tokyo 181-8588, Japan
}%

\author[0000-0002-7443-2215]{Yudai Suwa}
\affil{%
 Department of Earth Science and Astronomy, Graduate School of Arts and Sciences, The University of Tokyo, Tokyo 153-8902, Japan
}%

\author[0000-0002-8816-6800]{John P. Hughes}
\affil{%
 Department of Physics and Astronomy, Rutgers University, 136 Frelinghuysen Road, Piscataway, NJ 08854-8019, USA
}%

\author[0000-0003-1518-2188]{Hiroyuki Uchida}
\affil{%
 Department of Physics, Graduate School of Science, Kyoto University, Kitashirakawa Oiwake-cho, Sakyo-ku, Kyoto 606-8502, Japan
}%

\author[0009-0006-7889-6144]{Takuto Narita}
\affil{%
 Department of Physics, Graduate School of Science, Kyoto University, Kitashirakawa Oiwake-cho, Sakyo-ku, Kyoto 606-8502, Japan
}%

\author[0000-0001-8338-502X]{Hideyuki Umeda}
\affil{Department of Astronomy, Graduate School of Science, University of Tokyo, 7-3-1 Hongo, Bunkyo-ku, Tokyo 113-0033, Japan}

\begin{abstract}
Simulations of pre-supernova evolution suggest that some intense shell burning can be so active that, in extreme cases, it can merge with the outer shell, changing the initial conditions for the supernova explosion. However, such violent activity in the interior of stars has been difficult to confirm from observations of stars. Here we propose that the elemental composition of O-rich ejecta in supernova remnants can be a tool to test for this kind of intense shell burning activity in the final stages of progenitor evolution. As an example, we discuss the origin of ``Mg-rich'' ejecta in the supernova remnant N49B. A high Mg/Ne mass ratio $\gtrsim 1$ suggests that the Ne- or O-burning shell has broken into or merged with the outer shell before the collapse. Such Mg-rich (or Ne-poor) ejecta has been identified in some other supernova remnants, supporting the idea that some destratification process, such as a shell merger, does indeed occur in the interiors of some massive stars, although they may not be the majority. Our results suggest that X-ray observations of O-rich ejecta in core-collapse supernova remnants will be a unique tool to probe the shell burning activity during the final stage of a massive star's interior.

\end{abstract}


\section{Introduction} \label{sec:intro}
In the final days of a massive star's life, a series of shell burnings is known to be crucial for determining the internal structure of the star. The stellar structure affects the propagation of the supernova shock wave, therefore the activity of the burning shells has recently attracted much attention from the point of view of the explosion mechanism of core-collapse supernovae \citep[e.g.,][]{2011ApJ...730...70O,2013ApJ...762..126O,2014ApJ...783...10S,2016ApJ...821...38S,2018ApJ...860...93S,2016MNRAS.460..742M,2016MNRAS.460.2664S,2013ApJ...778L...7C,2015ApJ...799....5C,2020ApJ...890..127C,2021Natur.589...29B,2012ApJ...757...69U,2016ApJ...818..124E,2012ARNPS..62..407J}. However, both observation and theory have struggled to describe the drastic evolution in the stellar interior during this short period. 

One of the most important parameters to link the final stages of the pre-supernova evolution to the supernova mechanism would be ``compactness''. The compactness of the pre-supernova core, which is closely related to a series of shell-burning processes in the final stages, controls the post-bounce accretion evolution and its remnant mass \citep[e.g.,][]{2011ApJ...730...70O,2014ApJ...783...10S}. The compactness is defined as 
\begin{equation}
\xi_M = \frac{M/M_\odot}{R(M_{\rm bary} = M)/1000~{\rm km}},
\end{equation}
where $R(M_{\rm bary} = M)$ is the radial coordinate that encloses a baryonic mass of $M$ at the time of core bounce \citep[][]{2011ApJ...730...70O}. Several studies have suggested that a smaller compactness more easily leads to supernova explosions \citep[e.g.,][]{2012ApJ...757...69U,2015PASJ...67..107N,2015ApJ...801...90P,2016ApJ...821...38S}.
More recently, \cite{2016ApJ...818..124E} have discussed a new criterion based on two parameters, $M_4$ and $\mu_4$, which are the mass enclosed at the point where the entropy in the progenitor exceeds a value of 4 $k_{\rm B}$ baryon$^{-1}$ and the mass derivative at the same position, respectively. $M_4$ has been used to locate the steep density drop often associated with a strong O-burning shell in the progenitor star \citep[e.g.,][]{2007PhR...442..269W}. In this criterion, a high value of $M_4$ combined with a low value of $\mu_4$ is favorable for an explosion, because both a high accretion luminosity, which accounts for a large fraction of the neutrino luminosity of the proto-neutron star at the time of shock revival, and a low mass accretion rate outside the $M_4$ interface are realized. 

Turbulent flow in the Si/O-burning shell is also known to be another important effect affecting the evolution of the supernova shock wave \citep[e.g.,][]{2013ApJ...778L...7C,2015ApJ...799....5C,2020ApJ...890..127C}, because the development and strength of neutrino-driven convection in the gain layer increase with increasing magnitude of the accreting seed perturbations \citep[e.g.,][]{2006ApJ...652.1436F,2008A&A...477..931S}. For example, using three-dimensional core-collapse supernova simulations, \cite{2013ApJ...778L...7C} have shown that aspherical velocity perturbations can alter the post-bounce evolution and trigger an explosion in a model that does not explode without them. This is known as a ``perturbation-aided'' explosion. However, the activity in the Si/O shell of a massive star has been difficult to discuss observationally, and currently the discussion of pre-supernova asphericity is mainly based on simulations. 

\cite{2020ApJ...890...94Y} have performed a three-dimensional (3D) simulation of a non-rotating 18.88 $M_\odot$ supernova progenitor \citep[see also][]{2016ApJ...833..124M,2019ApJ...881...16Y,2021MNRAS.506L..20Y} and observed a violent shell merger event, where the interface between the O/Ne layer and the Si/O layer disappears during the evolution, and Si is mixed far out into the merged O/Ne shell. In the merging process, Ne entrained inward by convective downdrafts burns, resulting in lower Ne/O ratio in the 3D model compared to the 1D model at the time of collapse. In addition to changing the elemental distributions and fractions in the interior of a massive star, the shell merger also affects the convective seed perturbations and the compactness of the stellar core, which are dynamically important for the explosion.

As shown in the theoretical studies above, the activity of a series of shell burning events in the final stage of stellar evolution is a key to understanding the initial conditions for the supernova explosion. In particular, shell mergers would be a fascinating phenomenon, affecting both compactness and seed perturbations. We propose here that X-ray observations of O-rich ejecta in supernova remnants could provide a unique opportunity to study such activity around the O-rich layer during the pre-supernova evolution. The majority of the O-rich ejecta should have been synthesized during hydrostatic nucleosynthesis during stellar evolution, and its elemental composition is not so sensitive to explosive nucleosynthesis \citep[e.g.,][and see also Appendix \ref{sec:presn-sn}]{1995ApJS..101..181W,1996ApJ...460..408T,2002RvMP...74.1015W,2012PTEP.2012aA302U}. This means that by measuring the elemental composition of the O-rich ejecta, the theoretically predicted disappearance of the O/Ne layer can be evidence for the intense shell burning activity. Such an observation of supernova remnants provides a unique test that is not possible with any observations of stars. In general, C-shell burning produces an O/Ne-rich layer at the endpoint of stellar evolution. On the other hand, some type of destratification, such as a shell merger, has the potential to produce an O/Mg-rich (Ne-poor) layer through enhanced Ne burning. Indeed, some ``Mg-rich'' ejecta have been identified in some supernova remnants \citep[e.g., N49B and G284.3-1.8:][]{2003ApJ...592L..41P,2015ApJ...808L..19W}, the origin of which is still unclear. In this study, we discuss the origin of the Mg-rich ejecta in N49B, which is known to be the remnant of a core-collapse supernova in the Large Magellanic Cloud (LMC) \citep{2003ApJ...592L..41P,2017ApJ...834..189P}. We propose that the destratification during the stellar evolution can be a new solution to explain the origin of the Mg-rich (Ne-poor) ejecta in supernova remnants.

\begin{figure*}[t!]
 \begin{center}
 \includegraphics[bb=0 0 1292 586, width=16cm]{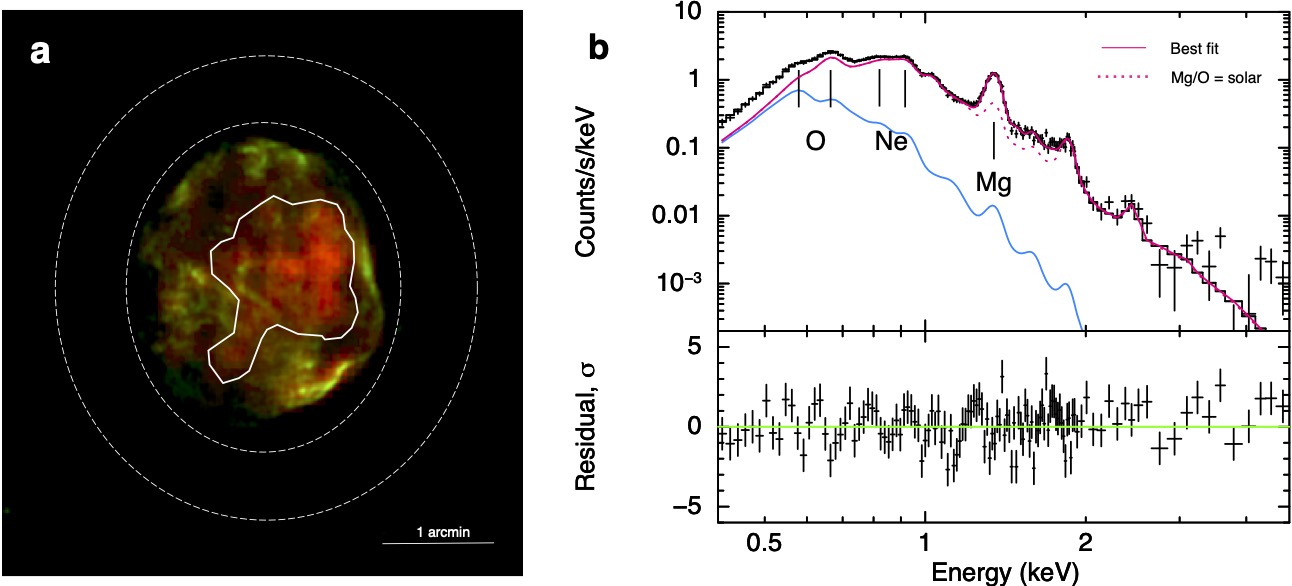}
 \end{center}
\caption{(a) Chandra X-ray image of N49B, combining images in green, and red made from energy bands of 0.5–10 keV (full band) and 1.32–1.42 keV (Mg K$\alpha$), respectively. The white contour shows the ejecta-dominanted region with strong Mg emissions for the spectral analysis. The dashed ellipses are the background region. (b) X-ray spectrum of the Mg-rich ejecta in N49B and its best-fit model (see solid lines). The blue and magenta curves represent the best-fit low- and high-temperature components, respectively. The dashed magenta line shows the plasma model that assumes the solar ratio between Mg and O.}
\label{fig:f1}
\end{figure*}

\begin{table}[]
    \centering
    \caption{The best-fit parameters for the spectral fitting. The errors show 90\% confidence level ($\Delta\chi^2$ = 2.7). The solar abundance in \cite{1989GeCoA..53..197A} is used.}
    \begin{tabular}{llc}
    \hline
        Components  &  Parameters                        & Best-fit values\\ \hline
        Absorption (phabs) &  $N_{\rm H}$ [10$^{21}$ cm$^{-2}$] & 2.0$^{+0.3}_{-0.1}$ \\
        ISM (nei)         & $kT_{\rm e}$ [keV] & 0.22$^{+0.11}_{-0.01}$\\
                          & Abundance           & 0.2 (fix) \\
                          & Redshift [10$^{-3}$] & 1.91$\pm$0.01\\
                          & EM [10$^{59}$ cm$^{-3}$] & 1.5$^{+0.1}_{-0.3}$\\
        Ejecta (vpshock)  & $kT_{\rm e}$ [keV] & 0.65$^{+0.01}_{-0.02}$\\
                          & (O/H)/(O/H)$_\odot$  & 0.64$^{+0.02}_{-0.04}$\\
                          & (Ne/H)/(Ne/H)$_\odot$  & 0.88$^{+0.03}_{-0.08}$\\
                          & (Mg/H)/(Mg/H)$_\odot$  & 2.30$^{+0.16}_{-0.10}$\\
                          & (Si/H)/(Si/H)$_\odot$  & 0.57$\pm$0.06\\
                          & (Fe/H)/(Fe/H)$_\odot$  & 0.35$^{+0.01}_{-0.04}$\\
                          & $n_{\rm e} t_{\rm low}$ [10$^{9}$ cm$^{-3}$ s]  & 1.9$^{+2.0}_{-1.1}$\\
                          & $n_{\rm e} t_{\rm high}$ [10$^{11}$ cm$^{-3}$ s] & 2.6$^{+0.3}_{-0.1}$\\
                          & Redshift [10$^{-3}$] & 1.91$\pm$0.01\\
                          & EM [10$^{58}$ cm$^{-3}$] & 7.9$^{+0.8}_{-0.3}$\\ \hline
                          & $\chi^2$/d.o.f          & 184.25/125 \\
    \hline
    \end{tabular}
    \label{tab:bestfit}
\end{table}

\begin{figure*}[t!]
 \begin{center}
  \includegraphics[bb=0 0 1425 991, width=16cm]{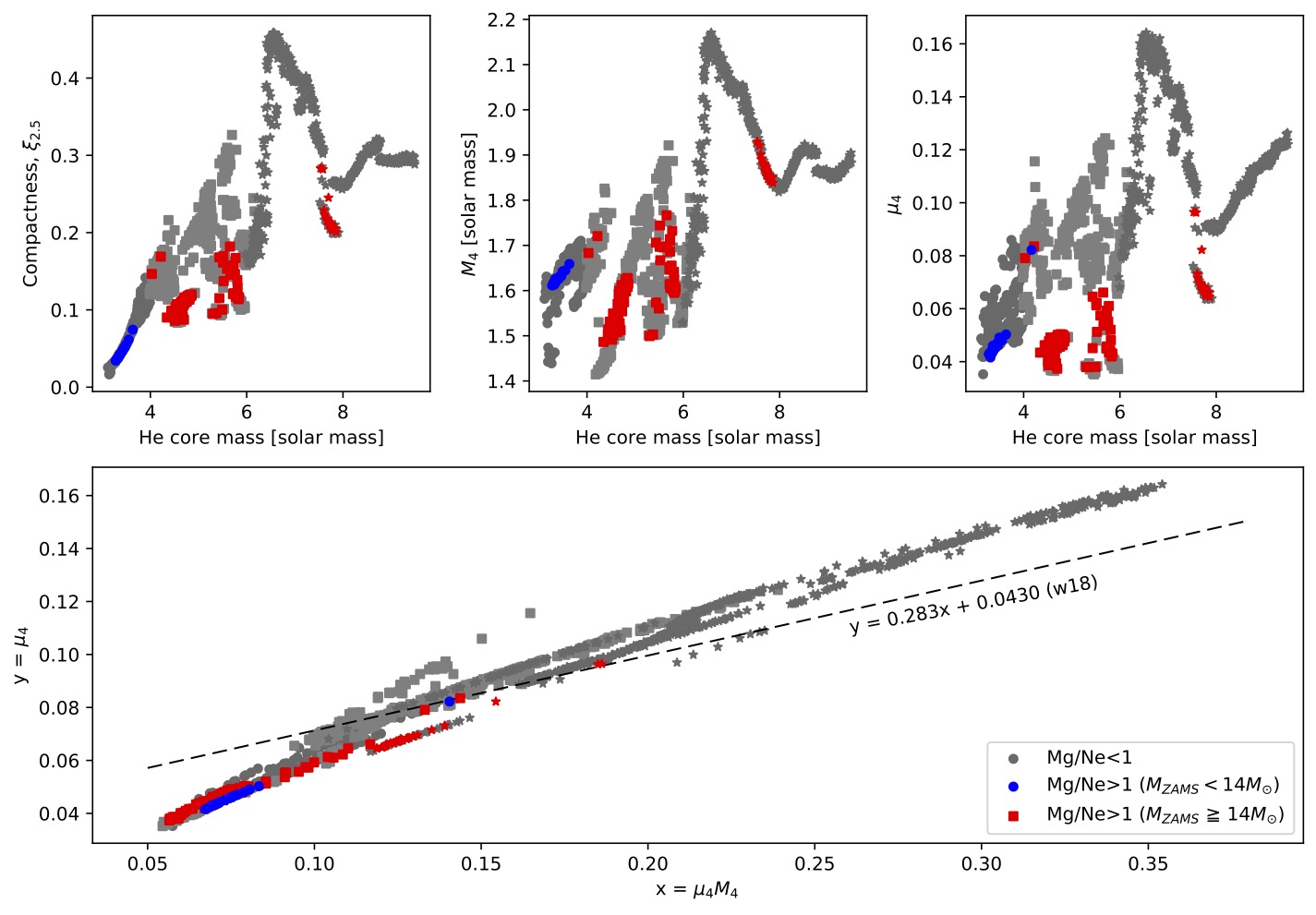}
 \end{center}
\caption{Top left: Compactness enclosing the innermost 2.5 $M_\odot$ ($\xi_{2.5}$) plotted as a function of He core mass. Top middle: $M_4$ plotted as a function of He core mass. Top right: $\mu_4$ plotted as a function of He core mass. Bottom: The relation between the Ertl parameters, $\mu_4M_4$ and $\mu_4$. The dashed line show a BH–SN separation curve for the w18.0 calibration shown in \cite{2016ApJ...818..124E}. In the gray data, the models have a low Mg/Ne mass ratio $<$ 1. The blue and red data show that the models have a high Mg/Ne mass ratio $>$ 1 with $M_{\rm ZAMS} < 14 M_\odot$ and $M_{\rm ZAMS} \geqq 14 M_\odot$, respectively. The circle, square and star data show the models in the mass ranges of $M_{\rm ZAMS} =$ 12--14 $M_\odot$, 14--19 $M_\odot$ and 19--27 $M_\odot$, respectively. All the parameters shown here were calculated by using the pre-supernova models provided by \cite{2018ApJ...860...93S}.}
\label{fig:compactness}
\end{figure*}

\section{Observation of Mg-rich Ejecta in N49B} \label{sec:obs}
Chandra ACIS-S has observed the LMC supernova remnant N49B on 2001 September 15 with a total exposure of 33.9 ks. We here reprocessed the event file of this observation (from level 1 to level 2) to remove pixel randomization and to correct for CCD charge transfer efficiencies using CIAO \citep{2006SPIE.6270E..1VF} version 4.13 and CalDB 4.9.5. The bad grades were filtered out, and good time intervals were reserved.

Figure \ref{fig:f1}a shows the Chandra image of N49B, with the Mg-K$\alpha$ emissions (1.32-1.42 keV band) highlighted in red. We extracted the X-ray spectra from the central ejecta-dominanted region. After subtracting the background spectrum from the surrounding empty sky (dashed ellipse), we fitted the X-ray spectrum with absorbed thermal plasma models (phabs$\times$(nei+vpshock) in Xspec) as shown in Figure \ref{fig:f1}b. The best-fit parameters are summarized in Table \ref{tab:bestfit}. Here, we assumed that N49B has two plasma components: a low-temperature ISM (nei: non-equilibrium ionization collisional plasma model) and a high-temperature Ejecta component (vpshock: plane-parallel shock plasma model). The abundance of the ISM component was fixed to be 0.2 to resemble the LMC abundance \citep[e.g.,][]{1998ApJ...505..732H}. For the Ejecta component, the abundances of S, Ar, Ca were linked to Si. The redshift parameters were linked between the ISM and Ejecta components, where the redshift velocity is estimated to be $\sim600$ km s$^{-1}$. This shift can be interpreted as the ACIS gain uncertainty of $\pm$0.3\%\footnote{https://cxc.harvard.edu/cal/summary/Calibration\_Status\_Report.html}, which is converted to the velocity uncertainty of $\approx \pm$900 km s$^{-1}$. The plasma parameters, $kT_{\rm e} \sim 0.7$ keV and $n_{\rm e}t \sim 3\times10^{11}$ cm$^{-3}$ s for the ejecta, agree well with the previous studies \citep[e.g.,][]{2003ApJ...592L..41P,2017ApJ...834..189P,2015ApJ...808...77U}.

We found a high Mg/Ne abundance ratio of 2.62$_{-0.26}^{+0.21}$, which is converted to a mass ratio of 0.97$_{-0.10}^{+0.08}$ assuming atomic weights of 24.3 and 20.2 for Mg and Ne, respectively. \cite{2003ApJ...592L..41P} and \cite{2017ApJ...834..189P} have already reported such ``Mg-rich'' ejecta with high Mg/Ne ratios in the central regions of the remnant. The spatially resolved analysis performed in \cite{2017ApJ...834..189P} suggests that the Mg/Ne abundance ratio varies from region to region in the range of $\sim$3--5, which is slightly larger than our estimate. \cite{2015ApJ...808...77U} have also reported a high Mg/Ne abundance ratio of $\sim$2.8 based on the Suzaku observation of N49B, which is in good agreement with our estimate. The difference amoung these Mg/Ne observations would come from different situations of spectral analysis (e.g., region selection, plasma modeling). However, even considering these observational uncertainties, there is a strong consensus on the high Mg/Ne mass ratio $\gtrsim$ 1 in the N49B ejecta. For example, the Mg/Ne mass ratio is 0.37--0.56 in the solar values \citep{1989GeCoA..53..197A,2009ARA&A..47..481A}, which is significantly lower than that in N49B. Since the solar abundances for Mg and Ne would be a sort of average of the contributions from massive stars, N49B with the high Mg/Ne ratio that deviates from the average could have experienced some peculiar stellar evolution or explosion.

\begin{figure*}[t!]
 \begin{center}
 \includegraphics[bb=0 0 1550 586, width=14cm]{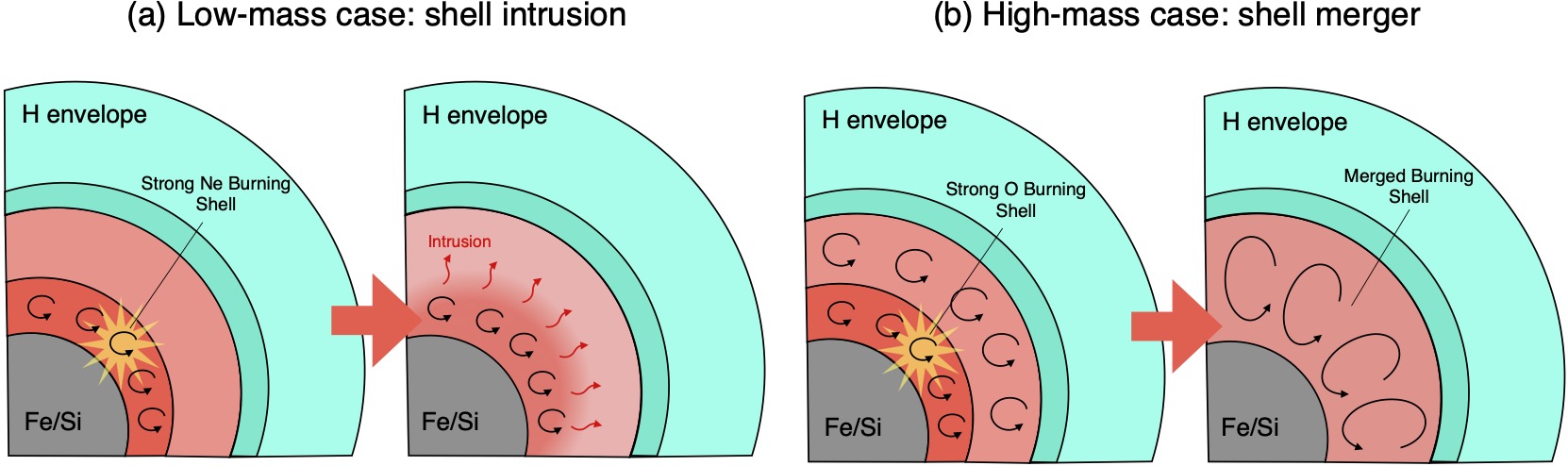}
 \end{center}
\caption{Schematic views of the destratification processes of shell-burning structures. (a) In the low-mass models with $M_{\rm He core} < 4~M_\odot$, strong Ne shell burning results in Ne-burning shell intrusion into the outer region. Here, the outer region is not a convective burning shell, thus there is no violent shell merger in this case. (b) In the high-mass models with $M_{\rm He core} > 4~M_\odot$, strong O shell burning produces a violent shell merger between the O and C burning shells. Both the destratification processes produce an Mg-rich composition in the O-rich layer by the combustion of Ne taken inward.}
\label{fig:interior_view}
\end{figure*}

\section{Discussion} \label{sec:discussion}
As already pointed out in the previous studies \citep{2003ApJ...592L..41P,2017ApJ...834..189P,2015ApJ...808...77U}, we found that the Mg-rich ejecta of N49B present a high Mg/Ne mass ratio of 0.97$^{+0.08}_{-0.10}$. However, it is still unclear how the Mg-rich ejecta were produced. In this section, we discuss the origin of the Mg-rich ejecta based on a comparison with theoretical models.

\subsection{Comparison with pre-supernova models} 
Because both the observed Mg and Ne in the N49B ejecta should have been synthesized in the O-rich layer of the progenitor star, we assume that the Mg/Ne ratio purely reflects the elemental composition in that layer. Of course, although the explosive nucleosynthesis could change the elemental composition, the explosive Ne burning region is too narrow to contribute to a significant change in the abundance of the entire O-rich layer (see Appendix \ref{sec:presn-sn} and Figure \ref{fig:presn-sn}). Therefore, we discuss here the cause of the high Mg/Ne ratio based on a comparison with that in pre-supernova models.

We analyzed 1,499 one-dimensional pre-supernova models (with the standard mass loss) provided by \cite{2018ApJ...860...93S} to search for destratification events that reproduce a high Mg/Ne ratio as observed in N49B. As pointed out by \cite{2020ApJ...890...94Y}, the disturbed shell activity is expected to reduce the amount of Ne in the O-rich layer as the Ne mixes inward and is burned. Not only in 3D simulations, some shell mergers has also been observed in 1D pre-supernova simulations \citep[e.g.,][]{1995ApJ...448..315W,2007PhR...442..269W,2002ApJ...576..323R,2007ApJ...671..821T,2014ApJ...783...10S}, where the compactness is tend to be small \citep[see section 3.4 in][]{2014ApJ...783...10S}. The interior of the star expands as nuclear burning is enhanced in the merged burning shell, which makes the compactness smaller. Therefore, some relationship between the Mg/Ne ratio and compactness is expected.

\begin{figure*}[t!]
 \begin{center}
 \includegraphics[bb=0 0 1060 938, width=16cm]{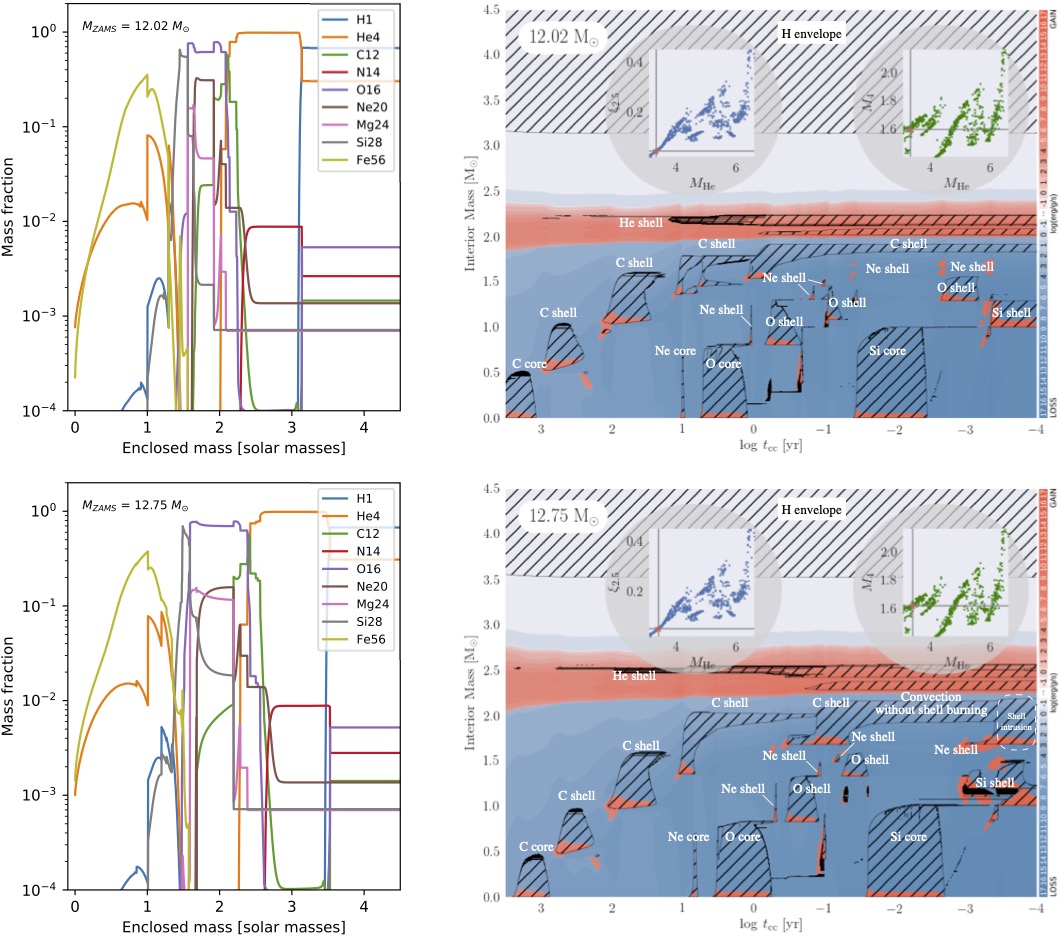}
 \end{center}
\caption{Top: The mass fraction profile and convective history of the 12.02 $M_\odot$ pre-supernova model, where the last Ne-burning shells are not convective. Bottom: Same as above, but for the 12.75 $M_\odot$ pre-supernova model, where the last Ne-burning shell is convective and breaks into the outer shell. The model data and convective plots are taken from \cite{VOEXDE_2018}.}
\label{fig:interior}
\end{figure*}

\begin{figure*}[t!]
 \begin{center}
 \includegraphics[bb=0 0 1060 934, width=16cm]{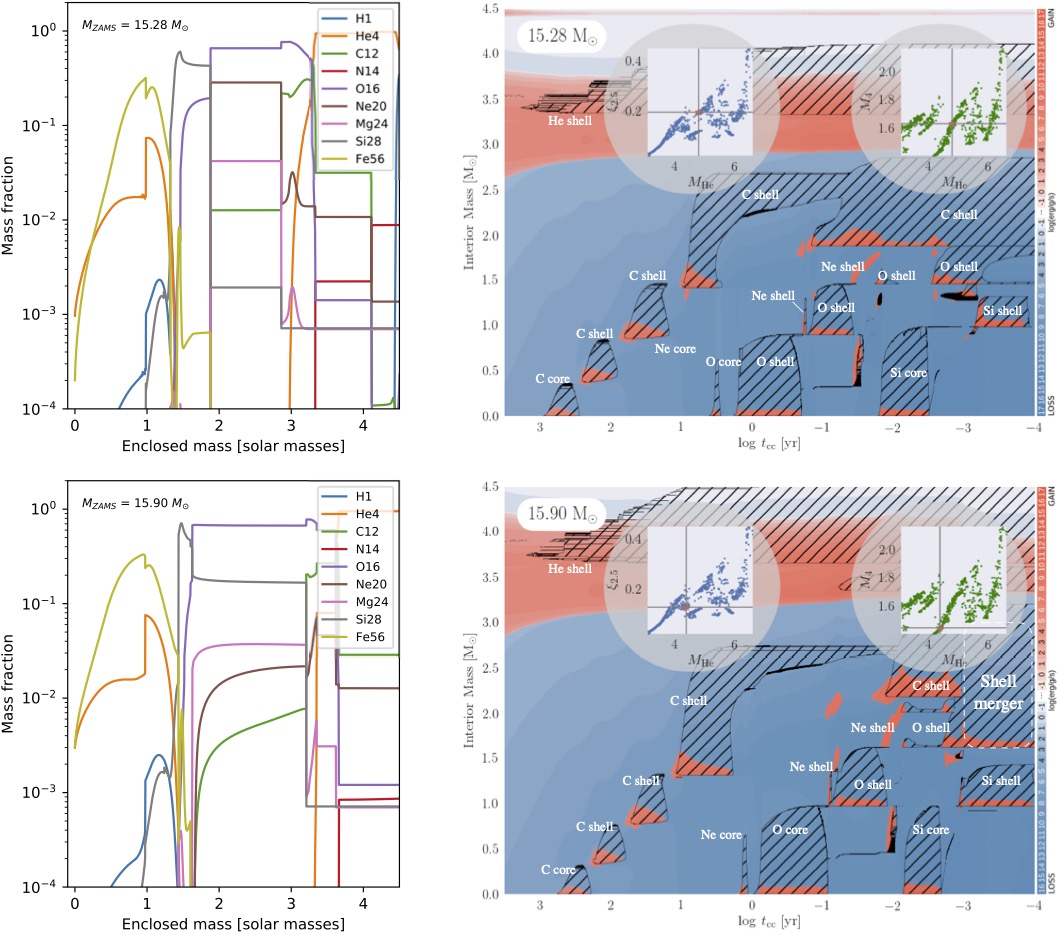}
 \end{center}
\caption{Same as Figure \ref{fig:interior}, but for the 15.28 $M_\odot$ and 15.90 $M_\odot$ pre-supernova models.}
\label{fig:interior2}
\end{figure*}

In Figure \ref{fig:compactness}, we clearly see that a model with a high Mg/Ne mass ratio $>$ 1 has a small compactness, $M_4$ and $\mu_4$ (see the detail discussions in section \ref{sec:exp}). Here we have calculated the Mg/Ne mass ratios in the O-rich layer of pre-supernova models, which has a high oxygen mass fraction $>$ 0.4. The Mg-rich models appear around regions where the compactness changes drastically, except for the models that have a low He-core mass below 4 $M_\odot$ ($M_{\rm ZAMS} \lesssim 14 M_\odot$).

In the low-mass models with $M_{\rm He core} < 4~M_\odot$, the Mg-rich cases are mainly the result of Ne-burning shell intrusion into the outer region (Figure \ref{fig:interior_view}(a)). As an example, Figure \ref{fig:interior} shows the mass fraction profiles and convective histories of the 12.02 $M_\odot$ and 12.75 $M_\odot$ pre-supernova models, where we have chosen the two similar-mass models that most easily compare the differences in elemental distributions. In the 12.02 $M_\odot$ model, the Ne-shell burning is not convective after the third O-shell burning (about 0.1 yr before the collapse), leaving a thin O/Mg/Si layer. On the other hand, in the 12.75 $M_\odot$ model, the Ne-shell burning becomes convective and breaks into the outer region shortly after $\sim$10$^{-3}$ yr before the collapse. The mixing process reduces the Ne fraction in the O-rich layer, resulting in a high Mg/Ne ratio in the O-rich layer.

In the high-mass models with $M_{\rm He core} > 4~M_\odot$, the Mg-rich models are considered as the result of shell mergers between the O- and C-burning shells (Figure \ref{fig:interior_view}(b)). In this case, the strong O-shell burning produces a violent merger of shells. As an example, Figure \ref{fig:interior2} shows the mass fraction profiles and convective histories of the 15.28 $M_\odot$ and 15.90 $M_\odot$ pre-supernova models, where we have chosen the two similar-mass models that most easily compare the differences in elemental distributions. In the 15.28 $M_\odot$ model, the fourth convective C-burning shell ignites about 0.1 years before the collapse, expanding into the region between 2 $M_\odot$ and 3 $M_\odot$ at the same time as the first O-shell ignites. This C-burning shell does not merge with the inner shell until the collapse, forming a flat O/Ne layer in the region between $\sim$2 $M_\odot$ and $\sim$3 $M_\odot$. In the 15.90 $M_\odot$ model, the last Ne- and O-burning shells expand and merge with the outer C-burning shell after $\sim$10$^{-3}$ yr before collapse, producing a large convective zone that expands between $\sim$1.6 $M_\odot$ and $\sim$3 $M_\odot$. Here, the mixing is more extensive than in the low-mass cases, and Ne is burned more efficiently, resulting in a higher abundance of Mg.

The most significant difference between the shell destruction processes for the low-mass and high-mass models is the degree of Si intrusion into the O-rich layer. In the former, the Ne-burning shell, where the Si fraction is low, breaks into the outer shell, producing a low Si/O ratio in the O-rich layer. On the other hand, in the latter, the O-burning shell, where the Si fraction is high, merges with the outer C-burning shell. As a result, the O/Ne layer disappears and a new O/Si layer is formed (see the mass region between 2 $M_\odot$ and 3 $M_\odot$ in the bottom of Figure \ref{fig:interior2}). In some extreme cases, the entire Si-rich layer is incorporated into the O-rich layer, which resembles the multi-dimensional simulation of \cite{2020ApJ...890...94Y}. Therefore, the Si/O mass ratio would also be important to characterize these shell-merger phenomena. 

\begin{figure}[t!]
 \begin{center}
 \includegraphics[bb=0 0 810 780, width=8cm]{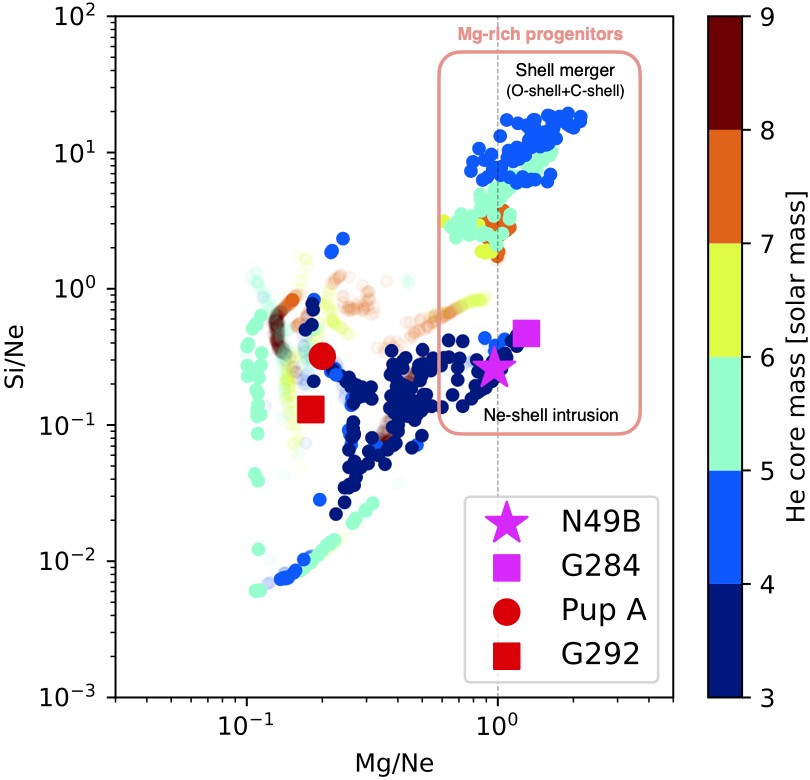}
 \end{center}
\caption{A scatter plot between the Mg/Ne and Si/Ne mass ratios derived from the pre-supernova models in \cite{2018ApJ...860...93S}. The magenta star shows the values of N49B obtained in this study. The magenta box shows the ratios in the western region of G284.3-1.8, which also shows a Mg-rich composition \citep{2015ApJ...808L..19W}. The red circle and box show the ratios in Puppis A and G292.0+1.8 \citep{2010ApJ...714.1725K,2014PASJ...66...64K}, which show the normal Ne-rich ejecta. The differences in He core mass are color-coded, where models that do not explode in Figure \ref{fig:compactness} are transparent in color. The dashed gray line shows the threshold for Mg-rich events (Mg/Ne mass ratio = 1). Shell mergers between the O- and C-burning shells, which occur in models with $M_{\rm Hecore} > 5 M_\odot$, result in a high Si/Ne ratio of $>$1.}
\label{fig:MgNe-SiNe}
\end{figure}

Indeed, the Mg/Ne vs. Si/Ne plot (Figure \ref{fig:MgNe-SiNe}) shows this tendency very well. We found that N49B has a low Si abundance, which currently supports the Ne-shell intrusion into the outer shell, which can be seen in the low-mass progenitors, as the origin of the Mg-rich ejecta in the remnant. However, there is a possibility that the Si abundance is diluted by mixing with the ISM gas, as seen in its sub-solar abundance. Also, in the low-mass progenitor case, the effects of explosive nucleosynthesis show unfavorable results for too large Mg/Ne mass ratios greater than 1, as seen in N49B (see Appendix \ref{sec:presn-sn} and Figure \ref{fig:presn-sn}). In addition, there would be uncertainties in pre-supernova models, thus we refrain from drawing strong conclusions on this topic. We note that the uncertainty in nuclear reaction rates such as $^{12}$C($\alpha$,$\gamma$)$^{16}$O also changes the values in Figure \ref{fig:MgNe-SiNe}.

\begin{table*}[]
    \begin{center}
    \caption{The number ratios between the O/Ne-rich and Mg-rich progenitors in \cite{2018ApJ...860...93S}, where ``Mg-rich'' means their Mg/Ne mass ratio is above 1.}\label{tab:rate}
    \begin{tabular}{cccccc}
    \hline
       12--13 $M_\odot$ & 13--14 $M_\odot$ & 14--15 $M_\odot$ & 15--16 $M_\odot$ & 16--17 $M_\odot$ & 17--18 $M_\odot$ \\
       0.19 & 0.02 & 0.07 & 0.56 & 0.07 & 0.12  \\\hline
       18--19 $M_\odot$ & 19--22 $M_\odot$ & 22--23 $M_\odot$ & 23--24 $M_\odot$ & 24--27 $M_\odot$ & 12--27 $M_\odot$\\
       0.15 & 0.00 & 0.13 & 0.14 & 0.00 & 0.10\\ \hline
    \end{tabular}
    \end{center}
\end{table*}

\subsection{Impact on supernova explosions}\label{sec:exp}
As shown in Figure \ref{fig:compactness}, the destratification processes that produce a high Mg/Ne ratio in the pre-supernova interior would also affect the explosion itself. Remarkably, almost all the Mg-rich models have the lowest $\mu_4$ values (see also Figure \ref{fig:compactness} top right). This is thought to be achieved by the expansion of the convection zone due to the intense shell-burning activity, which would lead to a low infall-mass rate into the stalled shock. Another interesting feature would be a relatively high $M_4$ value, which implies a high neutrino luminosity from its proto-neutron star, in the Mg-rich models with $M_{\rm Hecore} < 4 M_\odot$ (see blue points in Figure \ref{fig:compactness}). Both the low infall-mass rate and the high neutrino luminosity in the low-mass Mg-rich progenitors imply the easiest explosions among all the models, which may produce a more spherical and low-energy explosion. In the case of N49B, a relatively high explosion energy of $\sim$2--4$\times10^{51}$ erg was estimated \citep{1998ApJ...505..732H} and its shape has a smooth spherical form, which does not support all the features. While accurate estimates of explosion energy are always difficult in supernova remnant studies, future systematic discussions of this topic will be very interesting to understand the impact of intense shell activity on supernova explosions.

\subsection{Other candidates}
In many O-rich supernova remnants, their O/Ne layer appears to be intact without destratification. For example, famous O-rich supernova remnants such as 1E 0102.2--7219, G292.0+1.8, and Puppis A show high O/Ne ratios of $\sim$0.4 \citep[e.g., ][]{2004ApJ...605..230F,2014PASJ...66...64K,2010ApJ...714.1725K}, suggesting that destratification processes may not be a universal phenomenon in the evolution of massive stars (see red data points in Figure \ref{fig:MgNe-SiNe} for the normal O/Ne layer cases). Future systematic studies of the elemental composition of the O-rich ejecta in core-collapse supernova remnants will provide a fruitful discussion of the fraction of massive stars that undergo the destratification processes. 

Table \ref{tab:rate} shows the rate of destratificated Mg-rich events in the pre-supernova models in \cite{2018ApJ...860...93S}, which shows that 10\% of the pre-supernova models have a Mg-rich (Ne-poor) oxygen layer. In particular, the models in the mass ranges 12--13 $M_\odot$ and 15--16 $M_\odot$, where convection and shell burning conditions change significantly, have high Mg-rich event rates of 19\% and 56\%, respectively. In contrast to this, there are no Mg-rich events in the 19--22 $M_\odot$ and 24--27 $M_\odot$ mass regions where the compactness increases steadily. Given the initial mass function, the progenitors of $\lesssim$ 16 $M_\odot$ could account for the majority of destratification events. However, the model uncertainties are thought to be very large.

As pointed out in \cite{2018ApJ...860...93S}, the C- and O-convective shells are separated from coupling by only a single thin zone in some models, which may actually appear as other shell merger events. Therefore, differences in the treatment of multidimensional convection and overshoot may affect the appearance of shell mergers. This means that observational calibration will be important to understand the true rate of destratification events. As a first step in the observational approach to this topic, we discuss here two Mg-rich/Ne-poor candidates, G284.3-1.8 and Cassiopeia A, that are likely to have undergone the destratification during their pre-supernova evolution.

G284.3-1.8 is known as a Galactic supernova remnant hosting a high-mass X-ray binary (HMXB) in the central region. \cite{2015ApJ...808L..19W} have reported a high Mg/Ne abundance ratio of $\sim$3.5 in the western region, which is converted to a mass ratio of $\sim$1.3 using the solar abundance ratios in \cite{1989GeCoA..53..197A}. As can be seen in Figure \ref{fig:MgNe-SiNe}, the elemental composition of its ejecta is similar to that of N49B. The most significant difference from the N49B case is that only a portion of the western side has the Mg-rich composition. If the shell merger causes the Mg-rich ejecta, the Mg-rich composition is expected to see in the entire O-rich layer because the O-rich layer is well mixed in by convection. To produce the asymmetric distribution of Mg/Ne abundance in the remnant, some special shell destruction process may be needed, however we cannot conclude it from the current observed facts alone.

Cassiopeia A is one of the most well-studied core-collapse supernova remnants in our Galaxy. \cite{1996A&A...307L..41V} have reported significantly low Ne and Mg abundances in the ejecta of Cassiopeia A. The Ne/O mass ratio was estimated to be $\sim$0.01-0.02 \cite[see also][]{2012ApJ...746..130H}, suggesting the possibility of the Ne consumption by shell merger. On the other hand, the Mg/Ne mass ratio of $\sim$0.3 is not large, in contrast to the theoretical predictions for shell mergers with Ne depletion. Therefore, it cannot be explained by the destratification process alone, as investigated in this paper.

As discussed above, few candidates can currently be concluded to have undergone shell destruction like N49B; a comprehensive survey of O-rich remnants to show what fraction of such events exist would be beneficial to the discussion of the internal structure of progenitors of supernova remnants.

\section{Summary and Conclusion} \label{sec:summary}
A series of shell burnings is known to be crucial for determining the internal structure of the star, which is also dynamically important for the explosion itself. Therefore, the activity of the burning shells has recently attracted much attention from the point of view of the explosion mechanism of core-collapse supernovae. Simulations of pre-supernova evolution suggest that some strong shell burning can be so active that, in extreme cases, it can merge with the outer shells, changing the initial conditions for the supernova explosion. However, such activity in the interior of stars has been difficult to confirm from observations of stars. 

In this study, we demonstrated that the elemental composition of O-rich ejecta in supernova remnants can be a tool to test for this destratification phenomenon. As a first example, we studied the origin of ``Mg-rich'' ejecta in the supernova remnant N49B. A high Mg/Ne mass ratio $\gtrsim 1$ in its ejecta suggests that the Ne- or O-burning shell has broken into or merged with the outer shell. Such Mg-rich (or Ne-poor) ejecta have been identified in some other supernova remnants, supporting the idea that shell mergers do indeed occur in the interiors of some massive stars, although they may not be the majority. Our results suggest that X-ray observations of O-rich ejecta in core-collapse supernova remnants will be a unique tool to probe the activity during the final stage of a massive star's interior.

\begin{acknowledgments}
We thank Stanford E. Woosley and Alexander Heger for providing some convective plots for the progenitors with the ZAMS mass above 20 $M_\odot$ and useful comments. This work was supported by the Japan Society for the Promotion of Science (JSPS) KAKENHI grant No. JP19K14739 and JP23K13128.
\end{acknowledgments}

\appendix
\section{Effects of explosive nucleosynthesis}\label{sec:presn-sn}
\begin{figure*}[h!]
 \begin{center}
 \includegraphics[bb=0 0 1266 706, width=16cm]{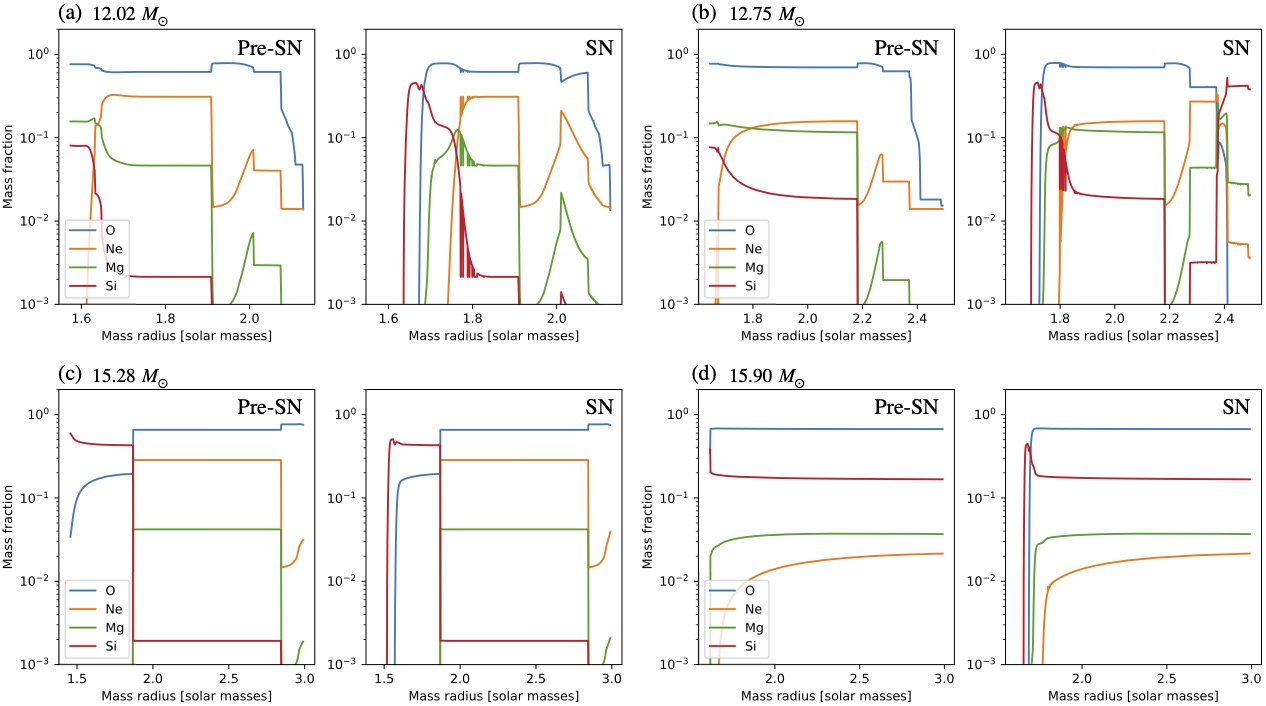}
 \end{center}
\caption{Comparison of the mass fraction profiles between pre-SN and SN models with the ZAMS mass of (a) 12.02 $M_\odot$, (b) 12.75 $M_\odot$, (c) 15.28 $M_\odot$ and (d) 15.90 $M_\odot$.}
\label{fig:presn-sn}
\end{figure*}

We have simulated 1D supernova explosions for the progenitors shown in Figure \ref{fig:interior} and \ref{fig:interior2} to check the effects of explosive nucleosynthesis. For numerical simulations, we employed the 1D Lagrangian hydrodynamic code in which neutrino heating and cooling terms are taken into account by the light-bulb approximation as described in \cite{2023arXiv230103610S} \citep[see also][]{2019MNRAS.483.3607S}. As a result, we found that the explosive nucleosynthesis does not change our conclusion, although the values change a little.

Figures \ref{fig:presn-sn} (a) and (b) show the results for the low-mass cases (12.02 $M_\odot$ and 12.75 $M_\odot$). If there is no destratification process (12.02 $M_\odot$), the O/Ne ratio in the O-rich layer shows little change, even though the inner part of the O-rich layer is scraped off by explosive nucleosynthesis. If there is the Ne-shell intrusion (12.75 $M_\odot$), the Mg-rich layer in the progenitor was burned by the explosion; thus, the contribution of Mg in the O-rich layer becomes smaller. This means that the Mg/Ne ratios in the low-mass cases were overestimated. For example, the Mg/Ne mass ratio in the O-rich layer of the 12.75 $M_\odot$ progenitor is 1.02 and the ratio becomes about 0.75 after the SN explosion. Thus, even if we consider the explosive nucleosynthesis, the ratio is more than twice higher than that without the destratification event of 0.32.

Figures \ref{fig:presn-sn} (c) and (d) show the results for the high-mass cases (15.28 $M_\odot$ and 15.90 $M_\odot$). In these cases, there are no major differences in the Mg/Ne ratio of the O-rich layer between the pre-SN and SN models, although the inner part of the O-rich layer is scraped off by explosive nucleosynthesis as in the low-mass cases. At least, the Mg/Ne mass ratio is greater than 1 in the shell-merger model (15.90 $M_\odot$) even if considering the effects of explosive nucleosynthesis.

\bibliography{sample631}{}
\bibliographystyle{aasjournal}

\end{document}